# Genome-wide Scan of Archaic Hominin Introgressions in Eurasians Reveals Complex Admixture History

**Authors:** Ya Hu,[1,3] Yi Wang,[1,3] Qiliang Ding,[1,3] Yungang He,[2] Minxian Wang,[2] Jiucun Wang,[1] Shuhua Xu,[2] and Li Jin[1,2]

[1] State Key Laboratory of Genetic Engineering and Ministry of Education Key Laboratory of Contemporary Anthropology, School of Life Sciences, Fudan University, Shanghai 200433, China.

[2] CAS-MPG Partner Institute for Computational Biology, Shanghai Institutes for Biological Sciences, Chinese Academy of Sciences, Shanghai 200031, China.

[3] These authors contributed equally to this work.

**Corresponding author:** Li Jin, lijin.fudan@gmail.com

**Abstract**

Introgressions from Neanderthals and Denisovans were detected in modern humans. Introgressions from other archaic hominins were also implicated, however, identification of which poses a great technical challenge. Here, we introduced an approach in identifying introgressions from all possible archaic hominins in Eurasian genomes, without referring to archaic hominin sequences. We focused on mutations emerged in archaic hominins after their divergence from modern humans (denoted as archaic-specific mutations), and identified introgressive segments which showed significant enrichment of archaic-specific mutations over the rest of the genome. Furthermore, boundaries of introgressions were identified using a dynamic programming approach to partition whole genome into segments which contained different levels of archaic-specific mutations. We found that detected introgressions shared more archaic-specific mutations with Altai Neanderthal than they shared with Denisovan, and 60.3% of archaic hominin introgressions were from Neanderthals. Furthermore, we detected more introgressions from two unknown archaic hominins whom diverged with modern humans approximately 859 and 3,464 thousand years ago. The latter unknown archaic hominin contributed to the genomes of the common ancestors of modern humans and Neanderthals. In total, archaic hominin introgressions comprised 2.4% of Eurasian genomes. Above results suggested a complex admixture history among hominins. The proposed approach could also facilitate admixture research across species.

**Introduction**

Ancestors of anatomically modern humans (AMH) and archaic hominins resided in the African continent after their divergence from chimpanzees approximately 5.6-7.6 million years ago (Mya) [1–7]. Previous reports suggested that archaic hominins migrated out of Africa into Eurasia at least 750,000 years before AMH appeared [8–12]. The earliest evidence of Neanderthals discovered in Eurasia was dated 400 thousand years ago (kya) [3]. The out-of-Africa theory suggests that the extant Eurasians originated from modern Africans, and they departed from Africa about 50 kya [10,11,13]. More recent studies showed that Neanderthals had admixed with Eurasians before their extinction ~30 kya [8,14], and 1-4% of the Eurasian genome is from Neanderthal introgression [8]. Another archaic hominin, Denisovan, had admixed with ancestors of Papuans and mainland Asians, but their contribution to Eurasians was much smaller than that from Neanderthals [9,12,15]. An unknown archaic hominin that diverged from the above hominins earlier than 1 Mya had contributed to Denisovans [15]. Archaic hominin introgression was detected in AMH genes, including *HLA* Class I, *STAT2,* and *HYAL2* [16–18]. However, a systematic effort to identify introgressed, non-AMH segments in modern human genomes is imperative for gaining a more detailed history of AMH, particularly its admixture with non-AMH hominins.

**Results**

**Identifying archaic hominin introgressions from 1000 Genomes Project Data**

To investigate chromosomal segments that exhibited features of an archaic hominin segment (AHS), we focused on introgressions from archaic hominins in extant Eurasian genomes. The divergence of archaic hominins and AMH (e.g., ~800 kya for Neanderthals)[4,8,9] occurred much earlier than the admixture between archaic hominins and Eurasians (47-65 kya) [19]; therefore, it is possible to identify introgressed AHS in the genomes of AMH. The primary challenges are first, detecting reliable signs of archaic hominin introgression, and second, locating the proper boundaries of AHS.

To address the first challenge, we used mutations in archaic hominins that occurred after their divergence with Africans as informative ancestral markers for AHS. These mutations are mostly absent in the African genome, and their number is much higher on the AHS than in the remainder of the Eurasian genome. In practice, we defined the *E-allele* as an allele that is absent from the African genome, but could be observed on any Eurasian chromosome, within a single nucleotide polymorphism (SNP) [12,20,21]. The first source of *E-alleles* comprised the mutations on AHS that occurred after the divergence between archaic hominins and Africans. The second source comprised the mutations on AMH segments in Eurasians that occurred after the Eurasian-African divergence. The third source comprised the alleles that existed in ancestors of Africans and Eurasians, but were lost in African samples, due to either genetic drift in the African genome or a limited African sample size. We further defined the *E-allele* rate of a segment as the number of *E-alleles* on the segment divided by the number of SNPs detected in the region encompassing the segment. Simulation results suggested that the *E-allele* rate was linear to the divergence time between the segment and the African genome (see Material and Methods). We distinguished AHS from other parts of the genome by their high *E-allele* rate. After we identified all AHS, we found that few *E-alleles* on AHS were derived from the third source, based on our estimations (see Material and Methods, Table 1).

We addressed the second challenge by partitioning the whole chromosome into segments, where different *E-allele* rates were found on neighboring segments. We achieved this partitioning with a method based on a Hidden Markov Model (HMM) [22], and we applied this method to the 1000 Genomes Project Phase1 data [23].

**Origins of identified archaic hominin introgressions**

We observed three peaks in the distribution of *E-allele* rates for all segments (Figure 1A). The segments in each peak were termed class1, class2, and class3 segments, which corresponded to ascending *E-allele* rates (Table 1). The three peaks represented different population divergence times from the African genome ($T_{afr}$); the different divergence times suggested that different waves of separation occurred between hominins and Africans.

The class1 genomic segments comprise the majority of the whole genome (97.6%) and are genetically closest to the modern Africans (*E-allele* rate = 0.0014). Therefore, we reason that class1 was contributed by AMH of African origin [8–12]. Given that the *E-allele* rate is proportional to the $T_{afr}$ (see Material and Methods), we estimated that the $T_{afr}$ of class2 and class3 segments are 895 and 3,464 kya, respectively, and the $T_{afr}$ of class1 was set at 50 kya [10,11].

Because the $T_{afr}$ of class2 segments was close to the reported divergence time between Africans and the ancestors of Neanderthals and Denisovans, we reason that the Neanderthals or Denisovans or both might be the sources of class2 segments [8,9,12]. To explore the origin of class2, we evaluated the similarity, defined as the percentage of *E-alleles* on class2 segments that were shared between class2 segments and the Altai Neanderthal genome ($S_{nean}$) or between class2 segments and the Denisovan genome ($S_{deni}$) (Figure 2) [12,15]. We observed that class2 segments bore much higher similarity to Neanderthal than to Denisovan genomes. Furthermore, the $S_{nean}$ distribution showed two modules, where the larger was separated from the smaller at $S_{nean}$=22.4%. The larger module peaked at around $S_{nean}$=76.8%, but that module was not observed in the $S_{deni}$ distribution. This analysis suggested that some class2 segments were derived from introgressions of Neanderthals or Neanderthal-related populations (denoted as class2N segments, $S_{nean}$> 22.4%, Figure S1, Figure S2), and that the Denisovan contribution was relatively minor [9,12,15]. Both $S_{nean}$ and $S_{deni}$ showed peaks near zero, which suggested that a substantial amount of class2 segments were unrelated to known archaic hominins (denoted as class2X, $S_{nean}$≤ 22.4%, Figure S3). Given their large sizes and their divergence from the African genome, about 27.9% of class2X segments were unlikely to be derived from the African genome (Table 1) [19]. These results suggested that a substantial number of class2X segments were not attributable to either Africans or any known archaic hominins; thus, we hypothesized that one or more unidentified archaic hominins (referred to as hominin-X hereafter) had contributed to the AMH genome.

Class3 segments were those most distantly related to the African genome, with a divergence (3,464 kya) about the time of, or earlier than, the exodus of *Homo erectus* out of Africa [7]. The probability that a segment was not broken by recombination, assuming a hypothetical African origin (*ρ*-value in Table 1), was low in class3 segments; thus, it was unlikely that class3 segments were derived from the African genome (Table 1) [19].

Overall, 37.1% of class3 segments were adjacent to a class2N segment (denoted as class3N segments, Figure S2). Simulation results did not support the hypothesis that class3N and class2N introgressed independently into the Eurasian genome (see Material and Methods). A majority (86.5%) of the class3N segments bore high similarity (>80%) to the Neanderthal genome. Thus, our results suggested that the class3N segments of unknown archaic hominins introgressed into the Eurasian genome via the Neanderthal genome introgression, together with the class2N segments [8].

The class3 segments that were not adjacent to class2 segments (denoted as class3E segments) accounted for 0.35% of the genome. These segments were flanked by AMH segments with low *E-allele* rates and low similarity with the Neanderthal genome (Figure S4) [10,11]. Furthermore, they were significantly shorter than the class2N segments (1:10 on average, $p<2.2\times10^{-16}$), which suggested that class3E segments existed in modern humans earlier than the time that Neanderthal genomic segments entered the Eurasian genome [19]. These results indicate that class3E segments could also be derived from unknown archaic hominins, but probably not from Neanderthal introgression. Interestingly, 28.5% of class3E segments overlapped with at least one class3N segment; thus, those segments were unlikely to have been derived from independent origins ($p<2.2\times10^{-16}$).

We observed only one module (close to a similarity=1) (Figure S5) in the distribution of pairwise similarities among the overlapping segments of class3. This finding suggested that class3 segments could be attributed to a single archaic hominin, which we refer to as hominin-E. We hypothesized that four possible models might explain the introgression from hominin-E to modern humans and Neanderthals (see Discussion). The most parsimonious model held that a single introgression occurred from hominin-E to the ancestors of modern humans and Neanderthals. This explanation was consistent with the reported results that an archaic hominin ($T_{afr}>1$ Mya) contributed to the Denisovan genome [15].

To conclude, we identified genomic contributions from two unknown archaic hominins (hominin-X and hominin-E) in the genome of modern humans. This discovery provides a substantial addition to previously proposed hypotheses [8,9,12,15], and it suggests a complex genomic admixture history of hominins. Figure 3 briefly summarizes our hypotheses.

**Comparison with existing Neanderthal introgression maps**

To further validate the identified AHS, we compared Neanderthal introgressions (class2N and class3N segments) with two published Neanderthal introgression maps [21,24]. For two published Neanderthal introgressions, 87.9% of regions in Vernot et al. [24] overlapped with identified Neanderthal introgressions, and 86.8% of regions in Sankararaman et al. [21] overlapped with identified Neanderthal introgressions. For identified Neanderthal introgressions, 44.8% of them overlapped with regions published by Vernot et al., and 81.1% of them overlapped with regions published by Sankararaman et al.

We further compared $S_{nean}$ among class2N and class3N segments overlapped and not overlapped with published Neanderthal introgressions. [21,24] We found both of segments overlapped and not overlapped with published regions bore high similarity with Altai Neanderthal (Figure 4). This suggests that detected Neanderthal introgressions were reliable, and the HMM based method could detect most of published Neanderthal introgressions.

For regions covered by introgressions from hominin-X (class2X segments), only 18.8% of them overlapped with regions published by Vernot et al. [24] and 21.6% of them overlapped with regions published by Sankararaman et al. [21] Both class2X segments overlapped and not overlapped with published regions bore low similarity with Altai Neanderthal (Figure S6), suggesting class2X segments were unlikely from populations related to Altai Neanderthals.

**Discussion**

In this study, we proposed an approach for identifying genomic segments from archaic hominin introgression into Eurasian genomes. In simulations, this approach achieved a false positive rate (i.e., the proportion of segments with an *E-allele* rate comparable to that of class1 segments, but identified as class2 or class3 segments) close to zero, when we selected the proper criteria. The detection power was ≥80% for class2 segments longer than 43 kb and class3 segments longer than 6 kb. Simulations suggested that this approach could locate the proper AHS boundaries, with a standard deviation of 8 kb and 1.5 kb for class2 and class3 segments, respectively. Over 98% of class2 and class3 segments contained <5% of *E-alleles* from the third source [20], which suggested that these *E-alleles* had little effect on the false positive rate in identifying AHS.

Moreover, when we applied *D* statistics, developed in a previous study on Neanderthal research [8], we found that *D (African, Neanderthal, class2X, chimpanzee)* was 0.009 with a standard deviation of 0.004. This result suggested that hominin-X might have shared ancestry with Neanderthals, when they diverged from the Africans. Hominin-E genetically diverged from the Africans very early (~3 Mya). One possible source of hominin-E is *Homo erectus*, which migrated out of Africa at ~1.7 Mya [7]. Another possible source could be the australopithecines, which inhabited Africa; it was proposed that those hominins might have migrated out of Africa at ~3 Mya [7].

There are seven possible scenarios that might explain the history of genomic admixtures between hominin-E and other hominins (Figure S7A-G). Class3N segments were derived from the hominin-E introgression in Neanderthals, and class3E segments were derived from the hominin-E introgression in modern humans. Because class3N and class3E segments accounted for only a small proportion of the genome (0.04% and 0.35%, respectively), the probability that 28.5% of class3E segments might overlap with class3N segments was less than $2.2 \times 10^{-16}$, when they were derived from different origins.

Therefore, the observed overlapping class3N and class3E segments could be derived from a hominin-E introgression, which occurred before the divergence of modern humans and Neanderthals [8,9,12]. Thus, the scenarios described in Figures S9A, S9B, and S9C are unlikely, because they could not explain the overlapping regions between class3N and class3E segments. Because 32.1% of class3N did not overlap with class3E and 71.5% of class3E did not overlap with class3N, modern humans and Neanderthals might have received an additional hominin-E introgression after their divergence. Therefore, the scenarios in Figures S9E, S9F, S9G are possible. Furthermore, the scenario in Figure S7D is also possible, because the non-overlapping regions of class3N and class3E segments could be explained by genetic drift after the divergence of modern humans and Neanderthals.

## Material and Methods

### Data sources

Whole-genome sequencing data for the modern human genome was obtained from the 1000 Genomes Project Phase 1 [23], which contains low-coverage DNA sequences from 1,092 individuals from 14 populations. The populations include YRI (Yoruba in Ibadan, Nigeria), LWK (Luhya in Wubuye, Kenya), CEU (Utah Residents with Northern and Western European ancestry, also known as CEPH), FIN (Finnish in Finland), GBR (British in England and Scotland), TSI (Toscani in Italia), IBS (Iberian population in Spain), CHB (Han Chinese in Beijing, China), CHS (Southern Han Chinese), JPT (Japanese in Tokyo, Japan), ASW (Americans with African Ancestry in Southwest USA), MXL (Mexican Ancestry from Los Angeles USA), CLM (Colombians from Medellin, Columbia), and PUR (Puerto Ricans from Puerto Rico). We combined YRI and LWK to represent Africans, CEU, FIN, GBR, TSI, and IBS to represent Europeans, and CHB, CHS, and JPT to represent East Asians. We combined Europeans and East Asians to represent Eurasians. We used the African and Eurasian datasets in the following analyses.

For the archaic hominin genomes, we obtained a high-coverage sequence from an Altai Neanderthal, low-coverage sequences from three Vindija Neanderthals [8], and a high-coverage sequence from an Altai Denisovan [12]. We used the filtering process used in Denisovan research for the three Vindija Neanderthals, and we combined the reads of the three Vindija Neanderthals [8,9]. For the Altai Neanderthal and Denisovan genomes, we filtered out reads with low mapping quality (<37) and low nucleotide calling quality (<40). Reads with indels were also filtered out. We then removed the first nucleotide on both ends of each read [8,9,12]. For a given SNP, when the count of the second most frequent allele was no less than 5% of the sum of the first and second most frequent alleles, the SNP would be called a heterozygote. Otherwise, the SNP would be called a homozygote.

**Testing the linear relationship between the *E-allele* rate and $T_{afr}$ with simulations**

To test the hypothesis that the *E-allele* rate of a segment was linear to its divergence time with the African genome, we used `ms` software to perform simulations of different demographics [25]. The simulated datasets contained chromosomes from two populations (African and an assumed population *E*). Population *E* diverged with the African population at time $T_{afr}$. We assumed a constant effective population size (*Ne*) for the African population. When population *E* diverged with the African population, the *Ne* of population *E* was assumed to be 1/10 of the African *Ne* [26]. The *Ne* of population *E* recovered to the level of the African *Ne* in 100 generations after the divergence with Africans. The command line of each simulation was:

```
ms 371 1 -s 10000 -I 2 370 1 -n 1 1 -n 2 1 -en t₁ 2 0.1 -ej Tafr 2 1
```

In the command line, *-s* indicates that there were 10,000 SNPs in each simulation. *–I* indicates that there were 370 chromosomes from the African population (the number of African chromosomes in the 1000 Genomes Project, Phase 1) [23] and one chromosome from population *E*. *–n* indicates that present *Ne* of Africans and population *E* are both one population size unit (a population size unit is 10,000). *–en* indicates that *Ne* of population *E* dropped to 1/10 of Africans when they diverged from Africans, then at time $t_1$ the *Ne* of population *E* recovered to the same level as the *Ne* of the Africans. *–ej* indicates that at time $T_{afr}$ population *E* diverged from Africans.

For each $T_{afr}$, we performed 100 simulations. Then, we averaged the *E-allele* rate of the 100 simulations to determine the *E-allele* rate of the chromosome from population *E*. The results showed that the *E-allele* rate was approximately linear to $T_{afr}$ (Figure S8), with a $R^2$=0.9837 ($p<2\times10^{-16}$). The intercept was slightly biased from 0 ($p<0.0469$), which indicated that *E-alleles* from the third source were present (see Results).

**Algorithm for identifying archaic hominin introgressions**

In this algorithm, we defined the *E-allele* rate as $r = E/(E+A)$, where $E$ is the number of *E-alleles* and $A$ is number of non *E-alleles* in a genomic segment (limited to the SNPs detected in the 1000 Genomes Project Phase 1 data) [23]. Given the long divergence time between archaic hominins and Africans (~800 kya for Neanderthals) [8,9,12], the *E-allele* rate on AHS was expected to be much higher than that in the rest of the genome. An important challenge in identifying AHS was to locate their boundaries. We addressed this challenge by partitioning the Eurasian chromosomes into segments with different *E-allele* rates.

We first implemented a dynamic programming approach to partition the chromosome. Given a Eurasian chromosome with $M$ SNPs, our goal was to partition the chromosome into $N$ segments. We set a linear penalty, $\lambda$, on $N$ to control the total number of segments and remove noise. The penalized log-likelihood function of the partitioning model (*PL*) was computed as follows:

$$PL = \sum_{i=1}^{N} E_i \ln r_i + A_i \ln(1 - r_i) - \lambda N$$

In the above equation, $E_i$ and $A_i$ are the numbers of *E-alleles* and non *E-alleles* on segment $i$, respectively, and $r_i$ is the *E-allele* rate of segment $i$. The dynamic programming approach maximized the *PL* via boundary optimization. This approach partitioned the problem into recursive sub-problems.

We defined $s(j,k)$ as the log-likelihood of the segment that started on the $j^{\text{th}}$ allele and ended on the $k^{\text{th}}$ allele:

$$s(j,k) = E_{jk} \ln(\frac{E_{jk}}{E_{jk} + A_{jk}}) + A_{jk} \ln(\frac{A_{jk}}{E_{jk} + A_{jk}}),$$

where $E_{jk}$ and $A_{jk}$ are the numbers of *E-alleles* and non *E-alleles*, respectively, in the segment.

We also denoted *b(l)* as the optimal penalized likelihood of the sub-problem that spanned the first SNP to the $l^{\text{th}}$ SNP, and *B(l)* was the optimal left boundary of the last segment in the sub-problem that spanned the first SNP to the $l^{\text{th}}$ SNP. We computed these with the following recursive formulas:

$$b(l) = \begin{cases} 0, l = 0 \\ s(1,1) - \lambda, l = 1 \\ \max(s(j,l) + b(j-1) - \lambda, 1 < j \leq l), 1 < l \leq M \end{cases}$$

$$B(l) = \begin{cases} 0, l = 1 \\ \max_j(s(j,l) + b(j-1) - \lambda, 1 < j \leq l), 1 < l \leq M \end{cases}$$

Starting from *l*=1, we solved the sub-problems one by one, and finally solved the full problem when *l=M*. When we obtained *B(M)*, we had obtained all the optimized boundaries recursively.

We implemented the above approach on the Eurasian chromosomes in the 1000 Genomes Project Phase 1 data with $\lambda$=10. We plotted the distribution of the *E-allele* rates in partitioned segments with the R function `density` (Figure S9). In the distribution, there were two modules and a tail (on the right), suggesting that there could be three main classes of segments.

The dynamic programming approach was computationally intensive; therefore, we replaced it with an efficient Hidden Markov Model (HMM) based method [22] for partitioning the chromosome. We assumed three hidden statuses in the HMM-based method, and each hidden status represented the level of *E-allele* rate in one of the three classes of segments identified with the dynamic programming approach.

We assumed that a Eurasian chromosome contained *M* alleles on *M* SNPs. $F_m$ was the *E-allele* status indicator for the $m^{\text{th}}$ allele. $F_m$=0 indicated that the allele was observed in the African genome, and $F_m$=1 indicated that the allele was absent from the African genome. Thus, we transformed *M* alleles into a string of 0 and 1 values. Our goal was to partition the string into $N$ segments and label each segment with a hidden status. Given three statuses (three *E-allele* rates) and the penalty for a status transition, the partitioning and labeling problem could be solved with a classic Viterbi algorithm through dynamic programming. After the partitioning and labeling were complete, we improved and re-estimated the initial *E-allele* rate by simple counting. The partitioning, labeling, and re-estimation were iterated until we achieved convergence.

The pseudocode of the HMM-based method was as follows:

1. Assume one chromosome, a penalty for status transitions, and an initial *E-allele* rate for each hidden status.

2. Use the Viterbi algorithm to partition the chromosome and label the hidden status of each segment.

3. Re-estimate the *E-allele* rate for each hidden status.

4. Repeat steps 2 and 3 until convergence.

**False positives of HMM based method**

We implemented the HMM-based method on the 1000 Genomes Project Phase 1 data and obtained three classes of segments (see Appendix 1). We assumed that class1 segments (the lowest *E-allele* rate among the three classes) were derived from AMH (see Results) [8–12]. We evaluated the false positive rate of the HMM-based method by simulation. We generated 1,000 AMH chromosomes with lengths of 300 Mb. SNP positions were uniformly distributed, and the SNP rate was 0.0125 per bp. The *E-allele* rate was set to 0.0015 (as in the empirical dataset). We implemented the HMM-based method with a different penalty $\lambda$ on the simulated dataset and evaluated the false positive rate; i.e., the proportion of AMH segments that were mistakenly identified as class2 or class3 segments (Table S1).

The *E-allele* rates of class2 and class3 segments were considered to be 0.026 and 0.112, respectively (as in the empirical dataset). We set $\lambda$=10 in all the following analyses to minimize the false positive rate.

**Detection power of HMM based method**

We then evaluated the detection power (i.e., one minus the false negative rate) of the HMM-based method. The length ($L$) of a class2 or class3 segment was expressed in units of kb. To evaluate the detection power for class2 or class3 segments of a given $L$ ($l$-1<$L$≤$l$), we generated 1,000 chromosomes of 1 Mb each. The SNP positions were uniformly distributed, and the SNP rate of each chromosome was set to 0.0125 per bp. The *E-allele* rate of each chromosome was set to 0.0015 to mimic the AMH (i.e., class1) segments. A segment (class2 or class3) with a known length ($l$-1<$L$≤$l$) and a known *E-allele* rate was placed at the center of each chromosome. The *E-allele* rate of the segment was set to 0.026 for class2 or 0.112 for class3. We applied the HMM-based method to the simulated dataset, and evaluated the detection power and the root mean square error (RMSE) for segments of different lengths. The detection power, at $l$kb, was the percentage of class2 and class3 (Figure S10) segments detected with $l$-1<$L$≤$l$. We found that, for class2 segments, the detection power was greater than 80% for $L$ > 43 kb (Figure S10). For class3 segments, the detection power was greater than 80% for $L$ > 6 kb (Figure S10).

**Proportion of class2 and class3 segments**

We then estimated the proportions of class2 and class3 segments in the Eurasian genome. For class2 segments shorter than 150 kb and class3 segments shorter than 30 kb, we adjusted their proportions in the genome based on the detection power. The detection power was close to 100% for class2 segments of 150 kb and class3 segments of 30 kb  (Figure S10A, Figure S10C), and it was higher for longer class2 and class3 segments. Therefore, we set the detection power to 100% for class2 segments longer than 150 kb and class3 segments longer than 30 kb. The proportion of class2 segments in the Eurasian genome (after adjustment) was expressed as:

$$(\sum_{l=1}^{150} \frac{\sum_{j=1}^{m_l} L_j}{p_l} + \sum_{l=151}^{L_{max}} \sum_{j=1}^{m_l} L_j) \times \frac{1}{N_{Euro}} \times \frac{1}{N_{base}},$$

where $L$ was the length of a class2 or class3 segment, in units of kb. In the equation, $p_l$ was the detection power of class2 segments with $l-1 < L \leq l$ ($0 < l \leq 150$); $m_l$ was the number of segments with $l-1 < L \leq l$; $L_j$ was length of the $j^{th}$ segment with $l-1 < L \leq l$; $L_{max}$ was the maximum length of class2 segments; $N_{Euro}$ was the number of Eurasian chromosomes in the 1000 Genomes Phase 1 data; and $N_{base}$ was the number of non-N bases in the human reference genome (GRCh37). The proportion of class3 segments in the Eurasian genome (after adjustment) was expressed as:

$$(\sum_{l=1}^{30} \frac{\sum_{j=1}^{m_l} L_j}{p_l} + \sum_{l=31}^{L_{max}} \sum_{j=1}^{m_l} L_j) \times \frac{1}{N_{Euro}} \times \frac{1}{N_{base}}$$

In the equation, $p_l$ was the detection power of class3 segments with $l-1 < L \leq l$ ($0 < l \leq 30$); $m_l$ was the number of segments with $l-1 < L \leq l$; $L_j$ was length of the $j^{th}$ segment with $l-1 < L \leq l$; $L_{max}$ was the maximum length of class3 segments; and $N_{Euro}$ and $N_{base}$ are defined above.

**Estimating the proportion of *E-alleles* from the third source**

The third source of *E-alleles* comprised alleles derived from the ancestors of Africans and Eurasians that were observed in Eurasian samples, but were absent in African samples, due to genetic drift or limited sample size. Next, we estimated the expected number of *E-alleles* from this third source for each segment [20].

We considered samples in the 1000 Genomes Project Phase 1 as a single population. For each segment, we obtained a minor allele frequency distribution for all SNPs in the region encompassing the segment. From this distribution, we estimated the probability that a sample was a homozygote of the major allele on a SNP (denoted as $p_h$) in the region encompassing the segment. We then estimated the probability that all 185 African samples were homozygotes of the major allele on a SNP ($p_h^{185}$). Therefore, the expected number of third source *E-alleles* on the segment could be expressed as $N_{SNP} \times p_h^{185}$, where $N_{SNP}$ is the number of SNPs in the segment region [20]. We then estimated the proportion of third source *E-alleles* expected in the observed *E-alleles* (denoted $p_1$, Table 1).

**Estimating divergence between Africans and archaic hominin introgressions**

Class1 segments were identified as AMH segments; therefore, we assumed that the $T_{afr}$ (divergence time with Africans) was 50 kya (38~64kya) as previously reported [10,11]. We estimated the $T_{afr}$ of class2 and class3 segments by comparing their *E-allele* rates with that of class1 segments.

The *E-allele* rate of a segment was linear to the $T_{afr}$ of its origin population, and over 98% of segments contained <5% of *E-alleles* from the third source. Thus, we assumed that *E-alleles* from the third source would have little effect on the relationship between the *E-allele* rate and the $T_{afr}$, and the *E-allele* rate was assumed to be proportional to $T_{afr}$.

For each class, we obtained the *E-allele* rate distribution with the *R* function `density`. The *E-allele* rate with the highest density ($R_E$) represented the *E-allele* rate of the class. Therefore, the $T_{afr}$ of a class could be expressed as $R_{EC}/R_{E1} \times 50$ kya, where $R_{EC}$ was the $R_E$ of the given class, and $R_{E1}$ was the $R_E$ of class1 segments.

The $R_E$ of class1 segments was 0.0014. The $R_E$ values of class2 and class3 segments were 0.025 and 0.097, respectively. Therefore, the estimated $T_{afr}$ values of class2 and class3 segments were 895 kya and 3,464 kya, respectively.

**Ancestry inference for class2X segments**

Because it was unlikely that class2X segments were derived from Neanderthal or Denisovan introgression ($S_{nean}$ and $S_{deni}$ close to 0) [12,15], we tested whether they were derived from the modern African genome. Under the null hypothesis that class2X segments were derived from the African genome, a segment would exist in the African genome no later than the divergence time between the African genome and the segment ($t$). The probability that a segment in the African genome was not broken by recombination was expressed as $\rho = e^{-\theta \times t}$ [20], where $\theta$ was the genetic distance between the two boundaries of the segment [27], and $t$ was expressed in units of generations. With a given $t$, $\rho$ was the probability that the genetic distance between the two boundaries of a segment was larger than $\theta$. A segment with a high $\theta$ would lead to a low $\rho$ value, and the rejection of the null hypothesis. For class2 and class3 segments, we estimated $t$ as $t = R_{ES}/R_{E1} \times 50$ kya $\times 1/25$, where $R_{ES}$ was the *E-allele* rate of the given segment, and $R_{E1}$ was the $R_E$ of class1 segments (0.0014). We assumed 25 years per generation.

Segments with $\rho < 0.05$ after the Bonferroni correction (for the number of segments tested) were not considered to originate from the African genome, but from archaic hominin introgression.

**Ancestry inference of class3N segments**

Class3N segments were defined as class3 segments that were adjacent to at least one class2N segment. We used simulations to test whether class2N and class3N segments were derived from independent archaic hominin introgressions. We simulated 1,000 Eurasian chromosomes with 1 million SNPs on each chromosome. We randomly sampled 2,000 SNPs along the chromosome, to represent recombination hotspots. We assumed that recombination occurred only at hotspots after the introgression of class2 and class3 segments [28,29]. Class3 segments were randomly distributed on chromosomes at a proportion of 1% before class2 introgression. We assumed that the initial proportion of class2 segments was 5%, and the time of class2 introgression was 2,000 generations ago [19]. The simulation results showed that the proportion of class2 segments was 1.4% and the proportion of class3 segments was 1.1%. The percentage of class3 segments adjacent to class2 segments was zero. This suggested that the observed adjacency of class3N and class2N segments could not be explained by independent introgression.

**Acknowledgements**

We thank Drs. S. Yan, P. Hu and H. Zheng for technical assistance.

**Financial Disclosure**

This work was supported by grants from the National Science Foundation of China (31271338, 31330038, 30890034, 31171218) and the National Basic Research Program (2012CB944600). The funders had no role in study design, data collection and analysis, decision to publish, or preparation of the manuscript.

**Competing Interest**

The authors have declared that no competing interests exist.

**Abbreviations**

AHS, archaic hominin segment.

**Figure Legends**

**Figure 1. The distribution of *E-allele* rates for genomic segments identified by the HMM-based method.** The distribution was obtained with the R function, `density`. **(A)** The *E-allele* rates in all segments identified. **(B)** *E-allele* rates in class1 segments. The *E-allele* rate with the highest density ($R_E$) was 0.0014. **(C)** *E-allele* rates in class2 segments, with $R_E$=0.025. **(D)** *E-allele* rates in class3 segments, with $R_E$=0.097.

**Figure 2. Distribution of similarity (%) between class2 genomic segments and two archaic hominin genomes.** Similarity is defined as percentage of *E-alleles* on class2 segments that are shared between class2 segments and archaic hominin genomes. Similarities are shown for the archaic hominin genomes, Neanderthal ($S_{nean}$; solid line) and Denisovan ($S_{deni}$; dashed line).

**Figure 3. Introgression events among hominins.** Arrow *N* (~50 kya) indicates the previously reported introgression from Neanderthals to Eurasians [8,15]. Arrows *E*, $E_N$, $E_A$, and *X* are the introgressions proposed in this study. Arrow *E* (before ~400 kya) is the introgression inferred from observations, and arrow $E_A$ and $E_N$ are possible introgressions from hominin-E to modern humans and Neanderthals.

**Figure 4: Distribution of $S_{nean}$ for identified introgressions that overlapped and not overlapped with published Neanderthal introgressions** [21,24]**.** The solid black line represents class2N and class3N segments overlapped with regions published by Vernot et al. [24] The dashed black line represents class2N and class3N segments not overlapped with regions published by Vernot et al. The solid grey line represents class2N and class3N segments overlapped with regions published by Sankararaman et al. [21] The dashed grey line represents class2N and class3N segments not overlapped with regions published by Sankararaman et al.

**Tables**

**Table 1. Characterization of AHS.**

| | | Inferred origin | $T_{afr}$ (kya) | Median length (kb) | Proportion in genome | $R_E$ | $p_1$ | $p_2$ | $p_3$ |
|---|---|---|---|---|---|---|---|---|---|
| Class2 | Class2N | Neanderthals | 929 | 71.8 | 1.40% | 0.026 | 99.5% | 68.9% | 95.4% |
| | Class2X | Hominin-X | 859 | 52.5 | 0.60% | 0.024 | 98.5% | 27.9% | 96.4% |
| Class3 | Class3N | Hominin-E | 3,924 | 15.6 | 0.04% | 0.110 | 99.4% | 48.1% | 71.4% |
| | Class3E | | 3,164 | 7.1 | 0.35% | 0.089 | 98.7% | 44.8% | 84.1% |

$p_1$ is the percentage of segments with less than 5% of *E-alleles* from the third source. $p_1$ in class1 was 98.3%. The probability that a segment was not broken by recombination, assuming a hypothetical African origin, was expressed as $\rho = e^{-\theta \times t}$ [19]; where $\theta$ is the genetic distance between the two boundaries of the segment [27], and $t$ is the divergence time (in generations) between the segment and the African genome. $p_2$ is the percentage of segments with $\rho < 0.05$ (we applied Bonferroni correction for the number of segments tested) in the indicated class. $p_3$ is percentage of *E-alleles* that are different from the chimpanzee genome [21].

Figure 1

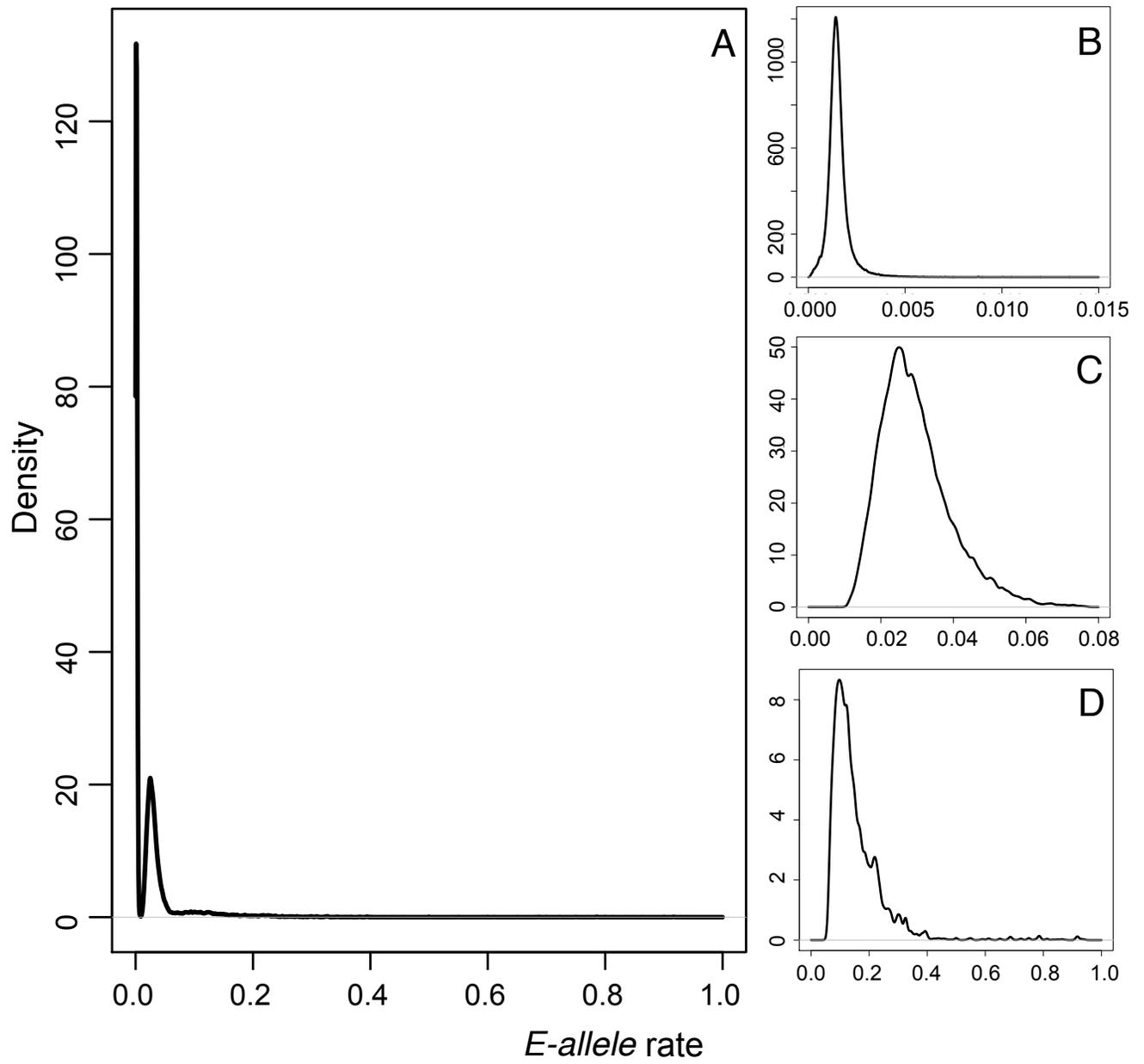

Figure 2

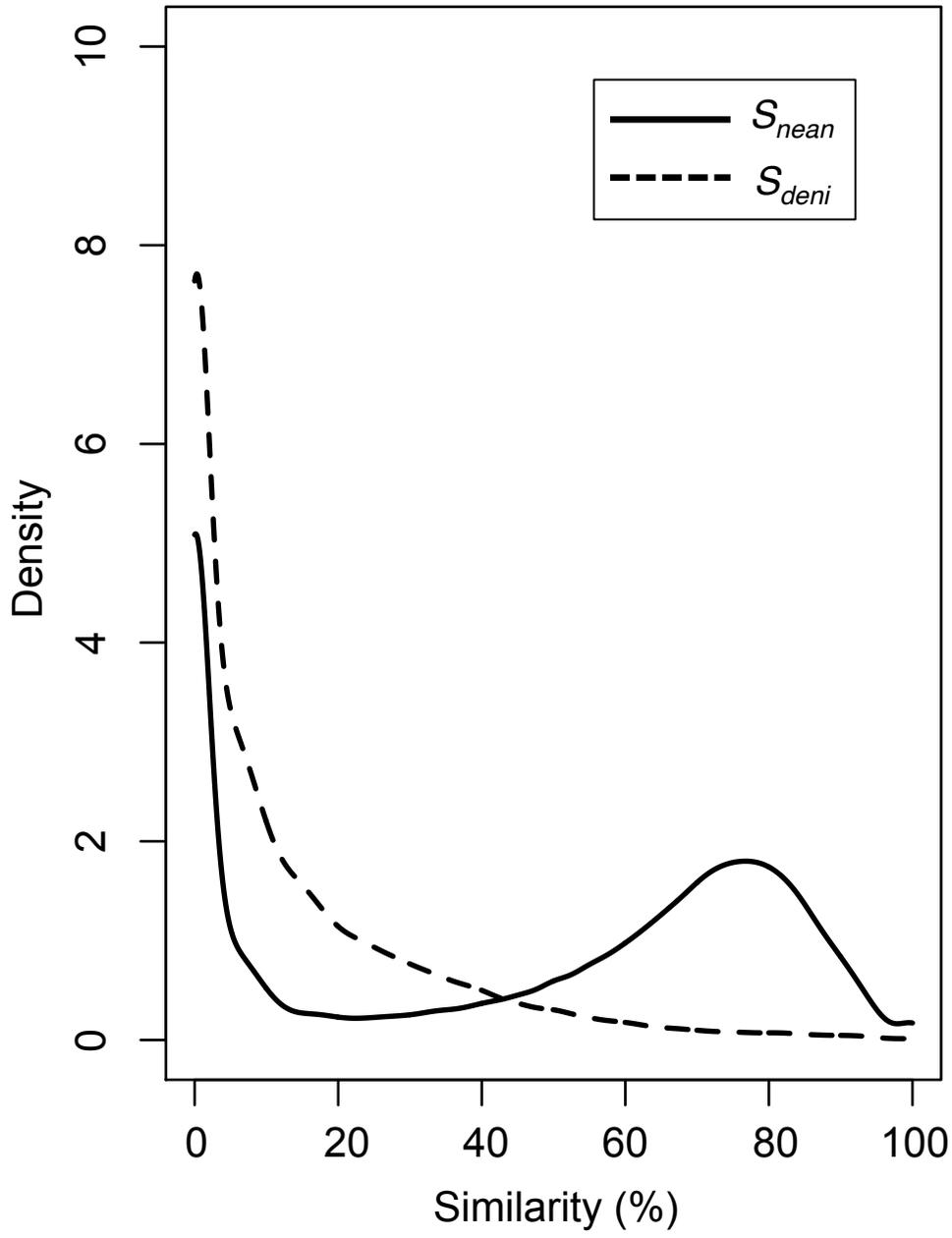

Figure 3

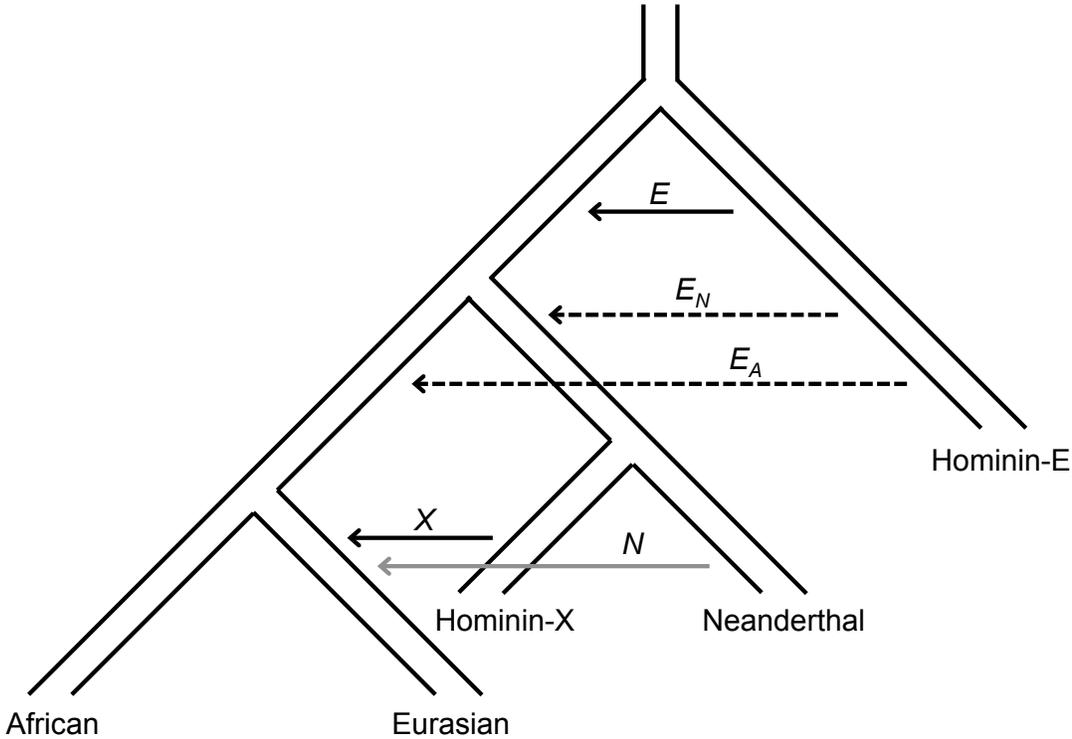

Figure 4

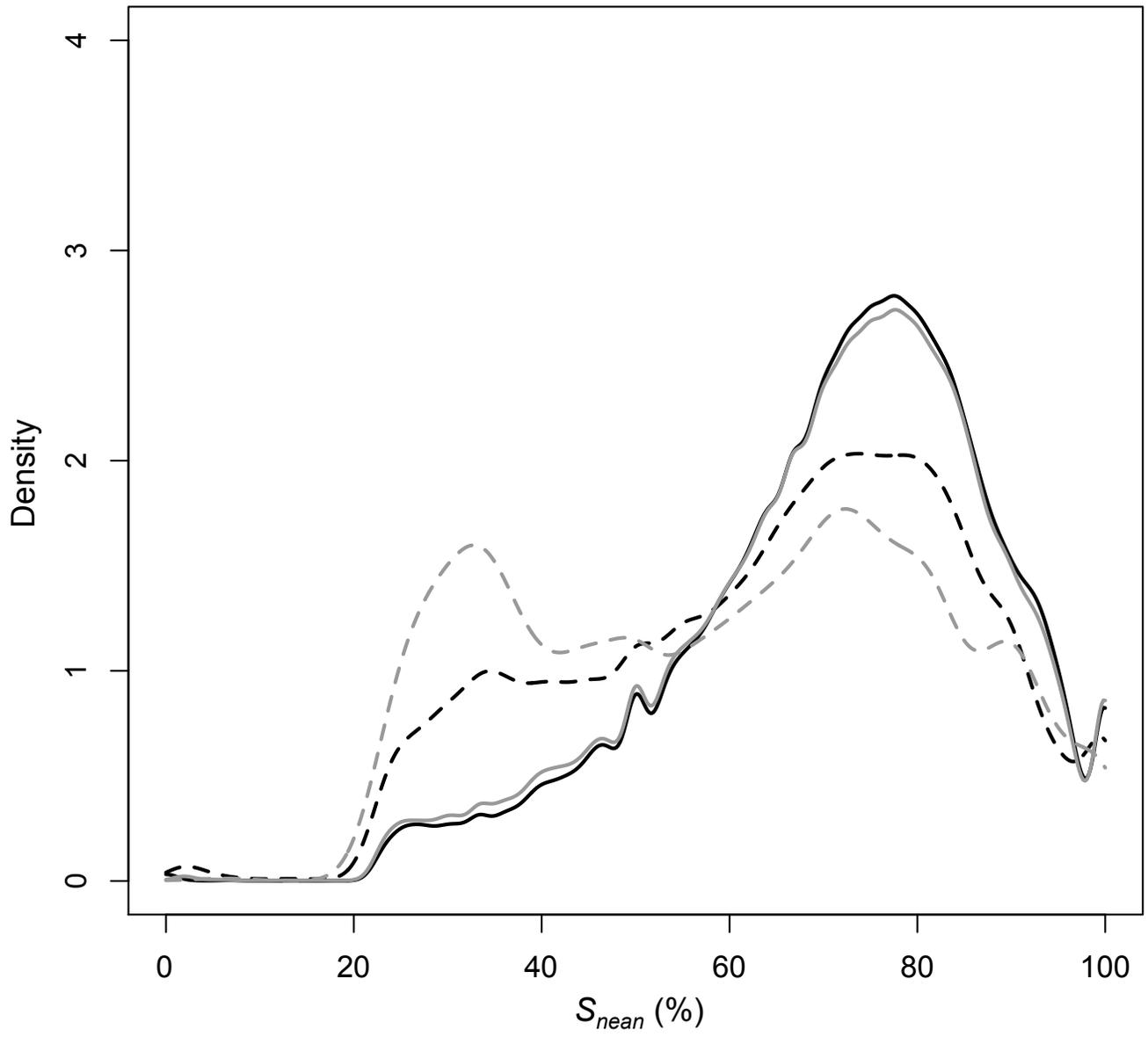

**Supporting Information**

**Figure S1. The distributions of similarity (%) between class2N segments and archaic hominin genomes.** Similarity is defined as percentage of *E-alleles* on class2N segments that are shared between class2N segments and archaic hominin genomes. The distributions were obtained with the R function `density`. (**A**) Similarity between class2N segments and the Altai Neanderthal genome. (**B**) Similarity between class2N segments and the Vindija Neanderthal genome. (**C**) Similarity between class2N segments and the Denisovan genome.

**Figure S2. The distributions of *E-allele* rates in class2 and class3 segments.** (**A**) *E-allele* rates in class2N segments. (**B**) *E-allele* rates in class2X segments. (**C**) *E-allele* rate in class3N segments. (**D**) *E-allele* rate in class3E segments.

**Figure S3. The distributions of similarity (%) between class2X segments and archaic hominin genomes.** Similarity is defined as percentage of *E-alleles* on class2X segments that are shared between class2X segments and archaic hominin genomes. (**A**) Similarity between class2X segments and the Altai Neanderthal genome. (**B**) Similarity between class2X segments and the Vindija Neanderthal genome. (**C**) Similarity between class2X segments and the Denisovan genome.

**Figure S4. The distributions of *E-allele* rates in segments that flanked class3E segments, and their similarity (%) to the Altai Neanderthal genome.** Similarity is defined as the percentage of *E-alleles* in the flanking segments that are shared with the Altai Neanderthal genome. (**A**) *E-allele* rate in the segments (50 kb) that flanked class3E segments. (**B**) Similarity between segments (50 kb) that flanked class3E segments and the Altai Neanderthal genome.

**Figure S5. The distribution of pairwise similarity (%) in the overlapping class3 segments.** We selected two overlapping class3 segments at a time, and computed similarity (%) as percentage of *E-alleles* that are shared between the two segments. The distribution includes all pairwise comparisons.

**Figure S6. Distribution of $S_{nean}$ for class2X segments that overlapped with and not overlapped with published Neanderthal introgressions** [21,24]**.** The solid black line represents class2X segments overlapped with regions published by Vernot et al. [24] The dashed black line represents segments not overlapped with regions published by Vernot et al. The solid grey line represents segments overlapped with regions published by Sankararaman et al. [21] The dashed grey line represents segments not overlapped with regions published by Sankararaman et al.

**Figure S7. Seven possible scenarios that could cause the observed admixture between hominin-E and other hominins.** (**A-G**) Branches indicate genomic divergence. Arrows indicate introgressions from hominin-E to other hominins.

**Figure S8. Linear relationship between the *E-allele* rate in the chromosome from population *E* and the $T_{afr}$ of population *E*.**

**Figure S9. The distribution of *E-allele* rates in segments partitioned with the dynamic programming approach.** The distribution was obtained with the R function `density`.

**Figure S10. Change of detection power with segment length in class2 and class3 segments.** (**A**) Relationship between the detection power and the length of class2 segments. (**B**) Relationship between the root mean square error (RMSE) and the length of class2 segments. (**C**) Relationship between the detection power and the length of class3 segments. (**D**) Relationship between the root mean square error (RMSE) and the length of class3 segments.

**Table S1. Penalties ($\lambda$) and false positive rates in simulated data.** We used $\lambda$=10 (in bold) for the penalty in all subsequent analyses.

Figure S1

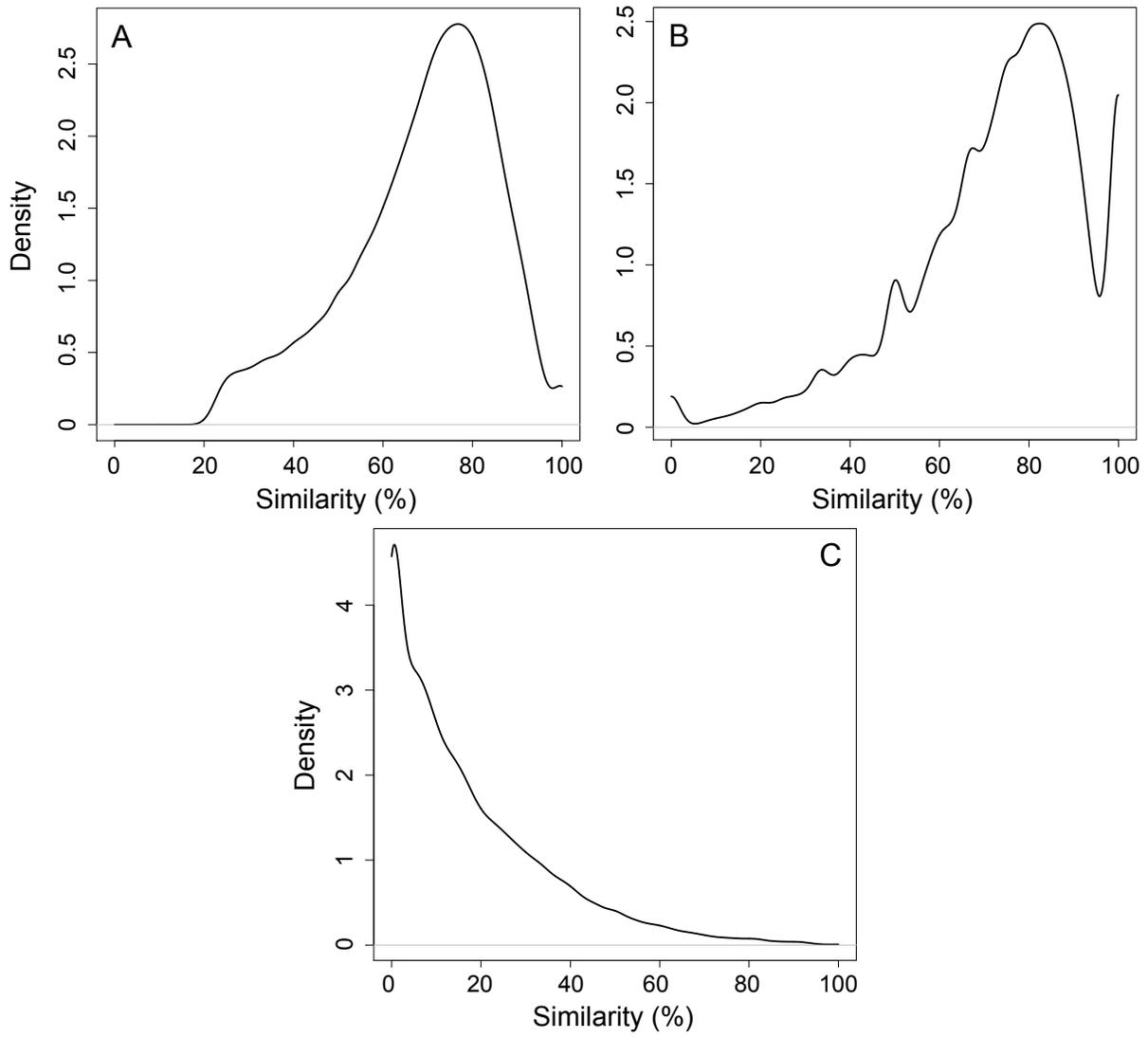

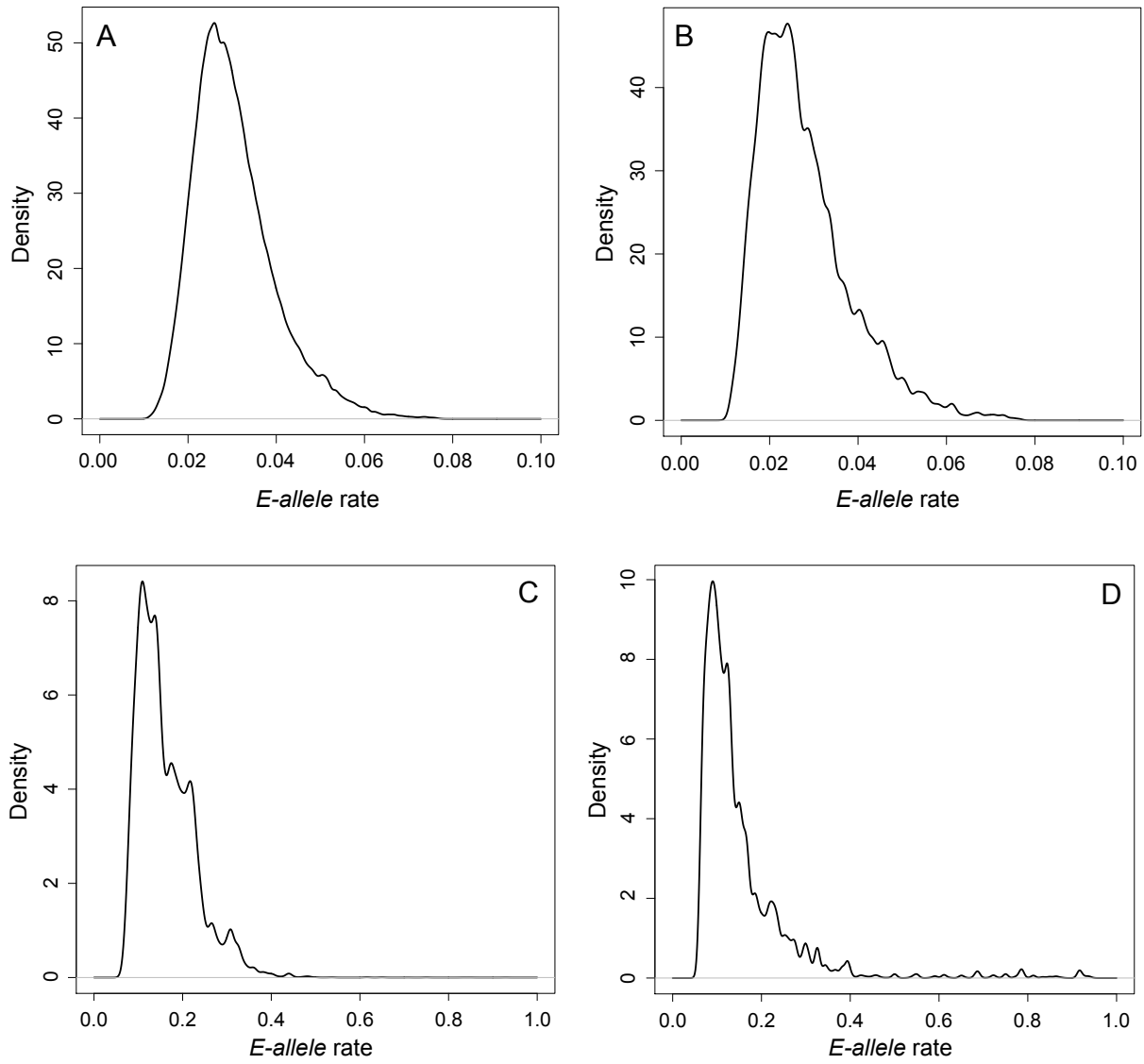

Figure S3

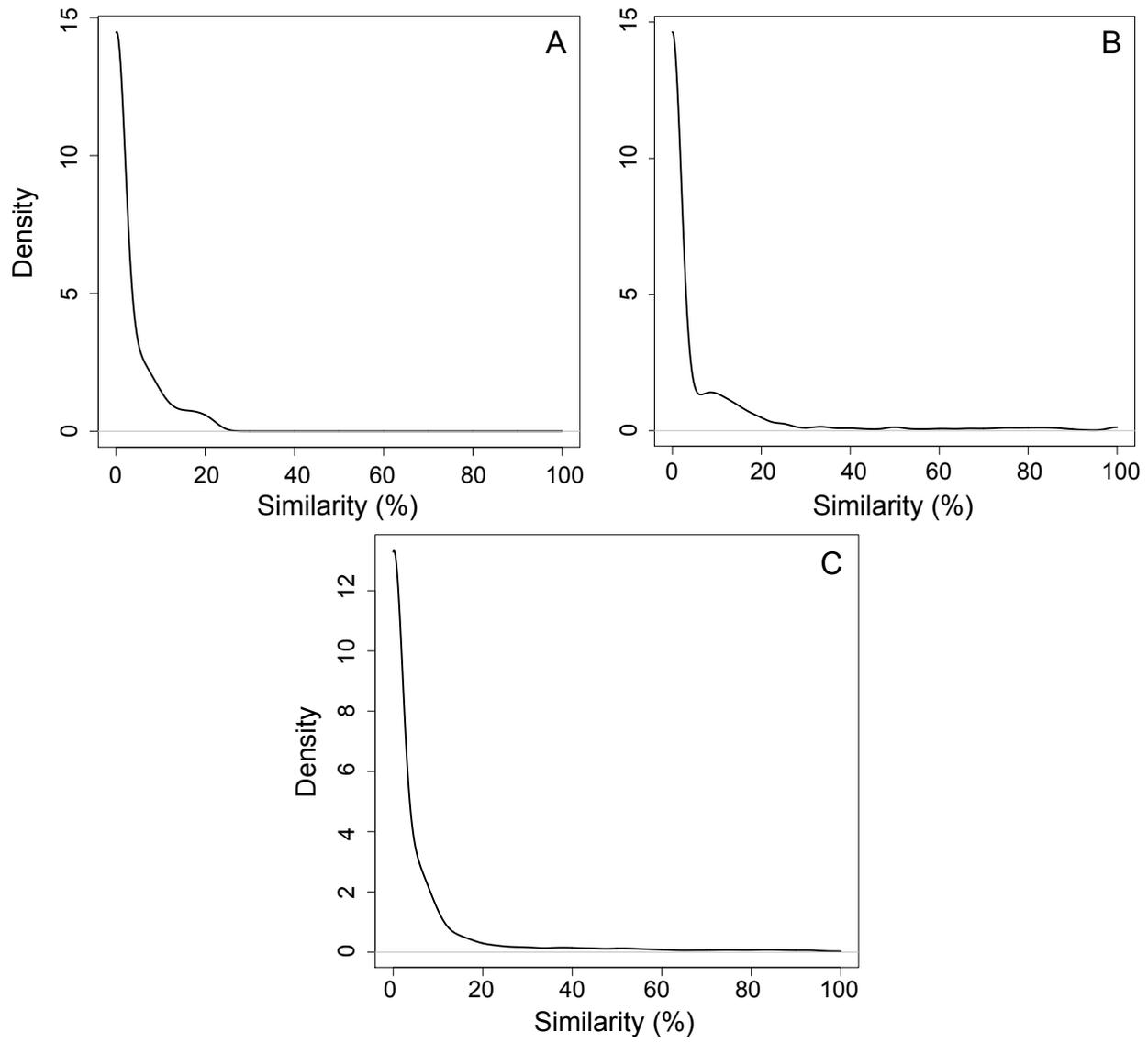

Figure S4

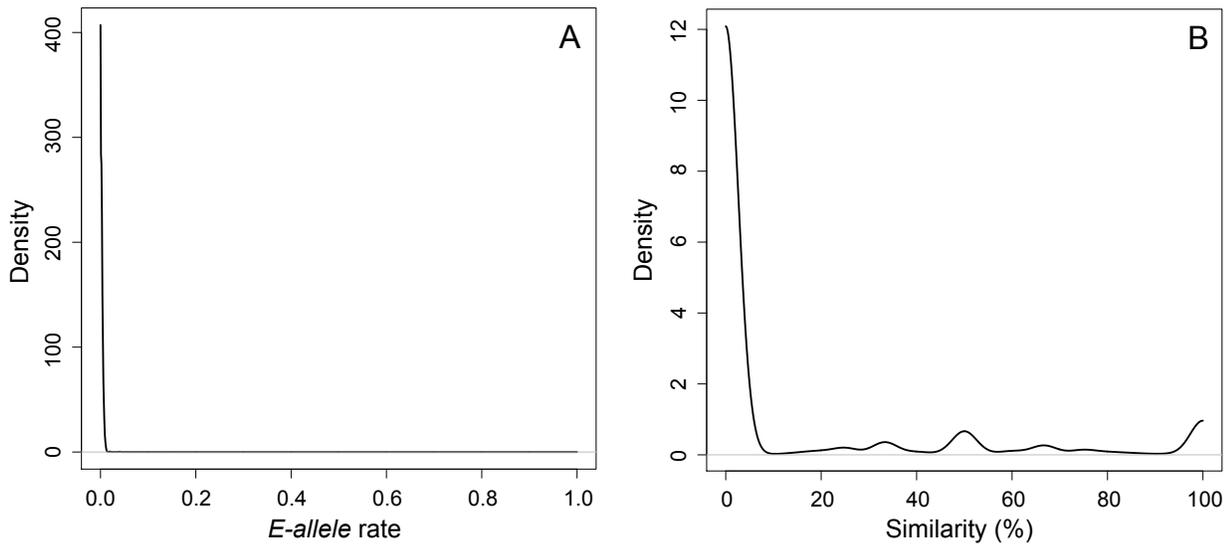

Figure S5

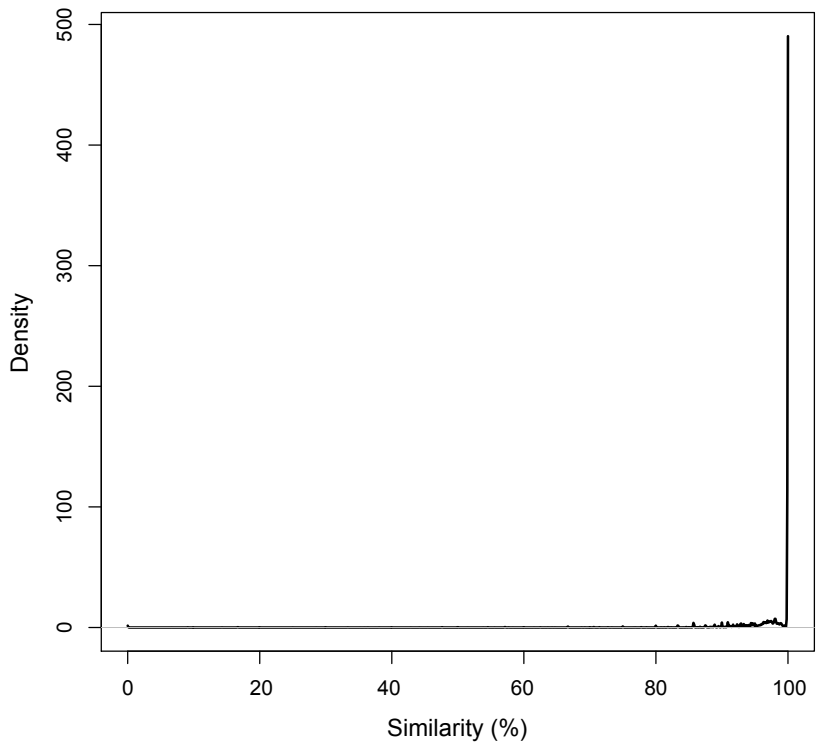

Figure S6

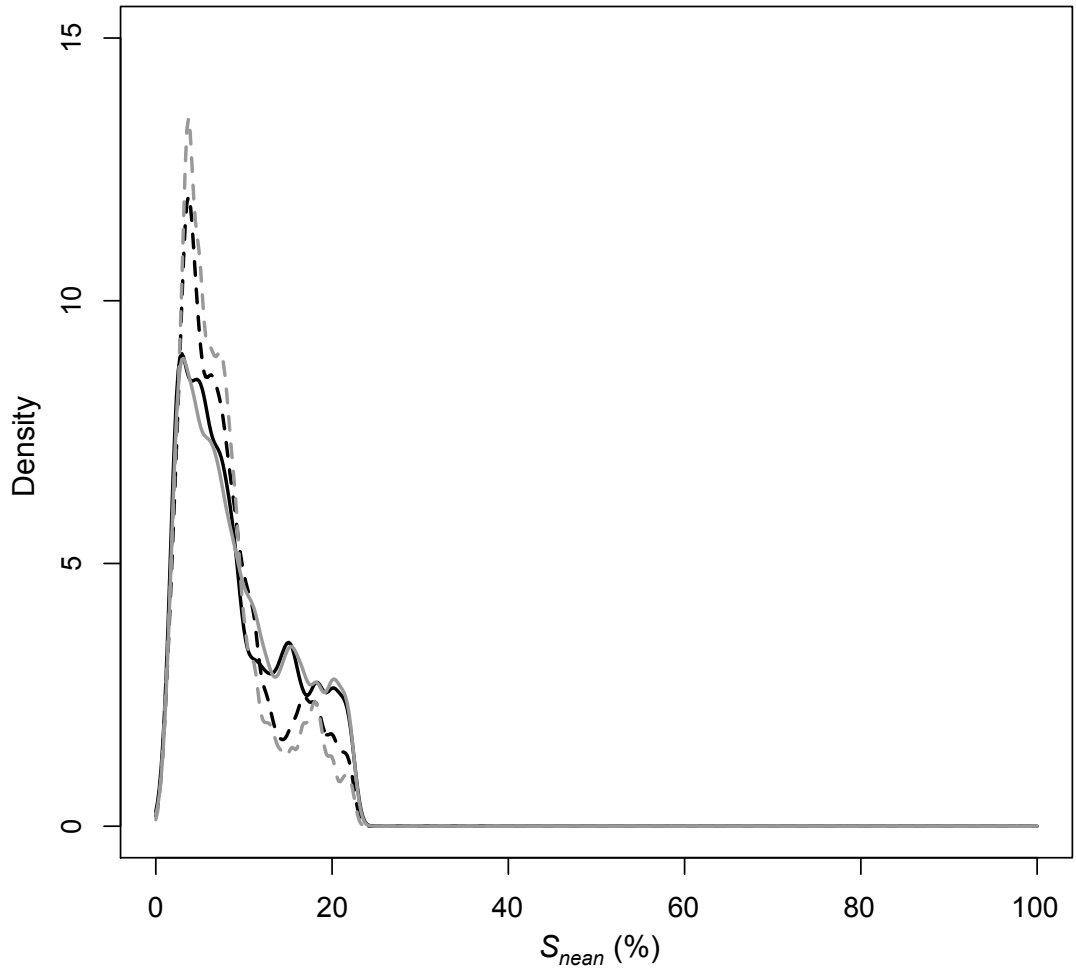

Figure S7

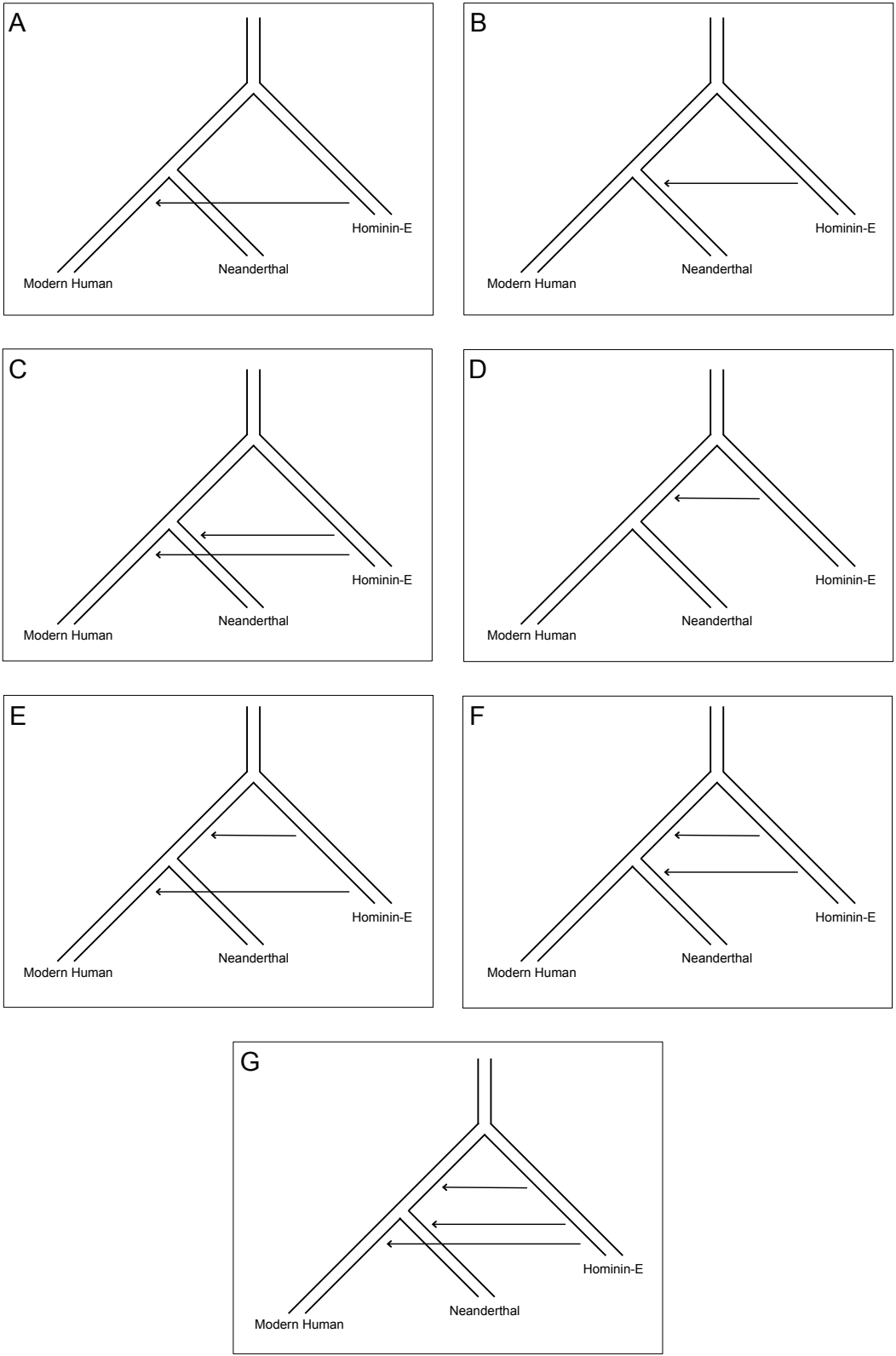

Figure S8

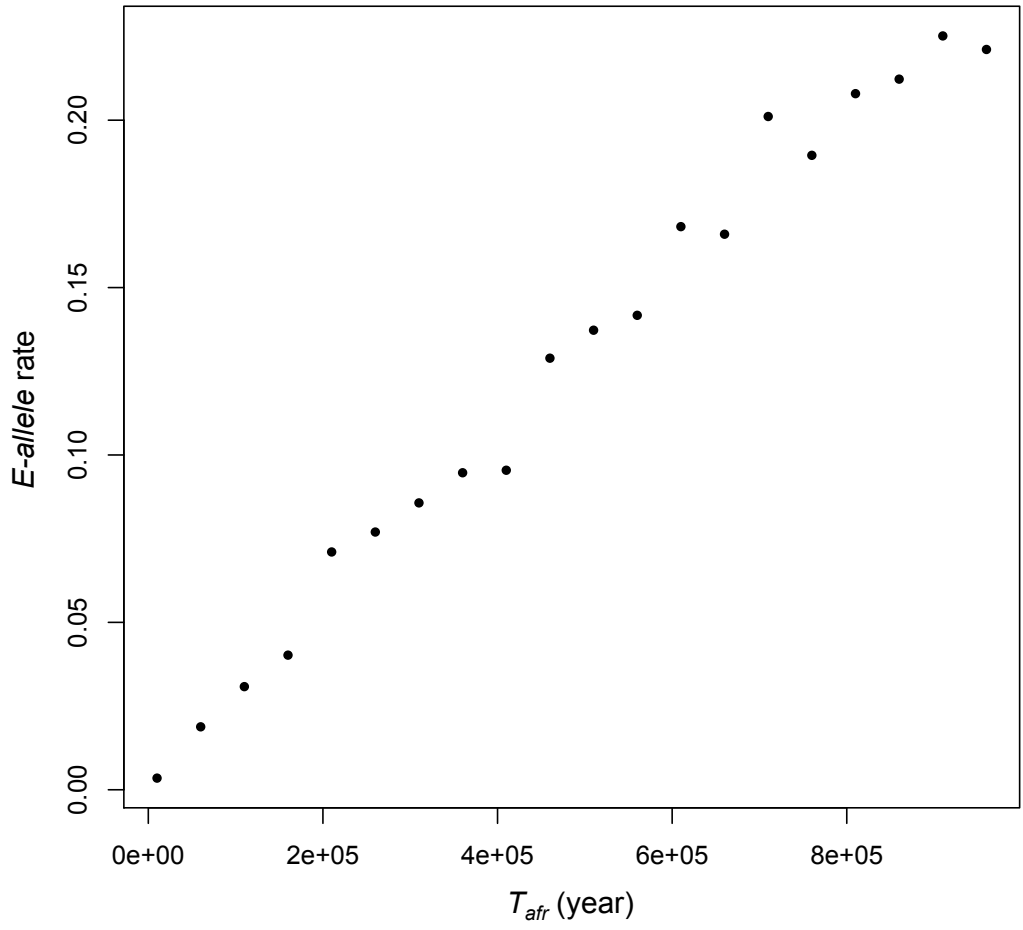

Figure S9

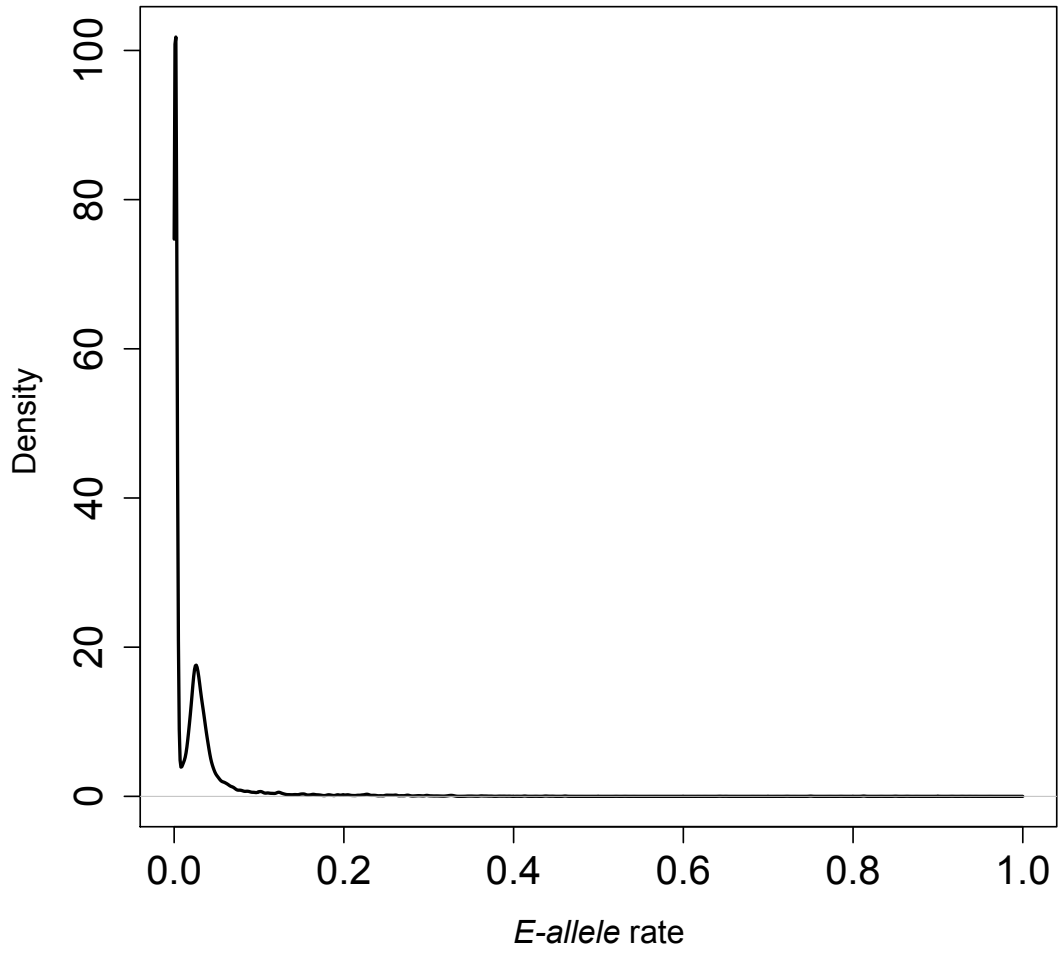

Figure S10

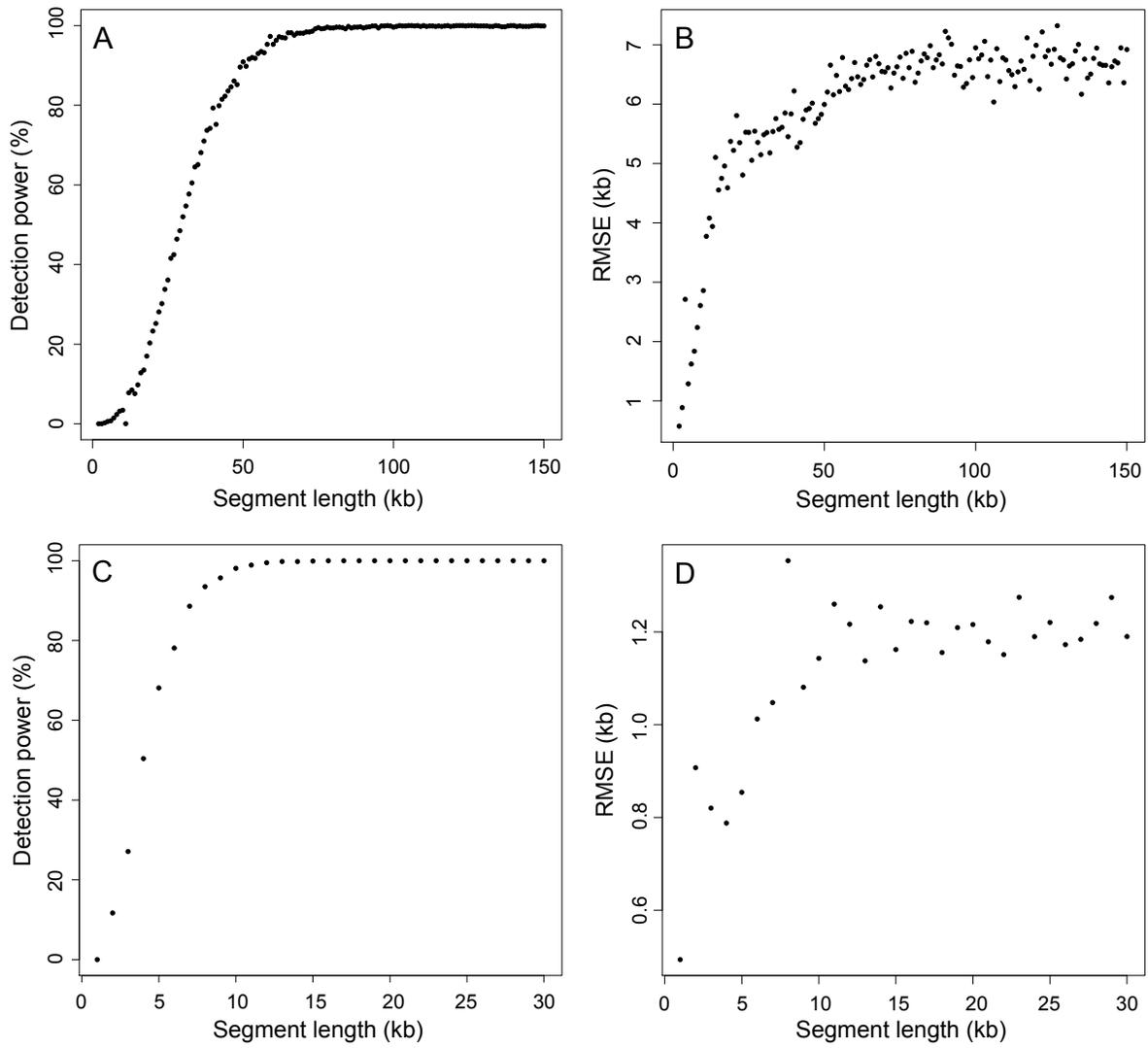

Table S1

| Penalty ($\lambda$) | False positive rate (%) |
|---|---|
| 6 | 9.3 |
| 7 | 1.1 |
| 8 | 0.2 |
| 9 | 0.0 |
| **10** | **0.0** |
| 11 | 0.0 |